\documentclass[envcountsame]{llncs}
\usepackage[T1]{fontenc}

\usepackage{amssymb,amsmath}
\usepackage{graphicx,tikz,hyperref,cleveref,framed,cite,xspace,mathtools,booktabs}
\usepackage{enumitem}

\usetikzlibrary{shapes,decorations,patterns,automata}

\newcommand{\xor}{\oplus}
\newcommand{\true}{\top}
\newcommand{\false}{\bot}

\newcommand{\framefont}[1]{\protect\ensuremath{\mathbf{#1}}\xspace}
\newcommand{\problemfont}[1]{\protect\ensuremath{\mathrm{#1}}\xspace}
\newcommand{\logicfont}[1]{\protect\ensuremath{\mathcal{#1}}\xspace}
\newcommand{\classfont}[1]{\protect\ensuremath{\mathrm{#1}}\xspace}
\newcommand{\setfont}[1]{\protect\ensuremath{\mathrm{#1}}\xspace}

\newcommand{\frameT}{\framefont{T}}
\newcommand{\frameSfour}{\framefont{S4}}
\newcommand{\frameSfive}{\framefont{S5}}

\newcommand{\PROP}{\setfont{PROP}}

\usepackage{etoolbox}
\usepackage{etoolbox}
\newcommand{\ML}[3]{%
  {\xor}\logicfont{ML}^{#1}_{%
    \ifstrequal{#2}{M}{M}{%
      \ifblank{#2}{}{{\{#2\}}}%
    }%
  }%
  \ifx#3\empty\else(#3)\fi%
}

\newcommand{\SAT}[4]{\framefont{#1}\text-{\xor}\problemfont{MSAT}^{#2}_{\{#3\}}\ifx#4\empty\else(#4)\fi}

\newcommand{\threeSAT}{\problemfont{3SAT}}

\renewcommand{\P}{\classfont{P}}
\newcommand{\NP}{\classfont{NP}}
\newcommand{\PSPACE}{\classfont{PSPACE}}

\usepackage[most]{tcolorbox}
\newtcolorbox{mytodo}[2][]{
  arc=5mm,
  lower separated=false,
  enhanced,
  breakable,
  title={#2}, 
  #1 
}

\usepackage{amsthm}
\usepackage{environ}
\NewEnviron{customproof}[1]{%
  \begin{proof}[Proof of #1]%
    \BODY%
    \hspace*{\fill}{\raisebox{0.5ex}{\scalebox{0.5}{\textsf{(#1)}}}}$\Box$\par%
  \end{proof}%
}

\author{Andreas Krebs\inst{1} \and Arne Meier\inst{2}}

\institute{Universit\"at T\"ubingen, Germany, \email{mail@krebs-net.de}
\and Leibniz Universit\"at Hannover, Germany, \email{meier@thi.uni-hannover.de}}

\authorrunning{A. Krebs and A. Meier}

\title{When Symmetry Yields NP-Hardness: Affine ML-SAT on S5 Frames}
\begin{document}
\maketitle

\begin{abstract}
Hemaspaandra~et~al.~[JCSS 2010] conjectured that satisfiability for multi-modal logic restricted to the connectives XOR and 1, over frame classes T, S4, and S5, is solvable in polynomial time. 
We refute this for S5 frames, by proving NP-hardness.
\end{abstract}

\section{Introduction}
Modal logic is an established formal concept that traces its roots back to the work of C.~I.~Lewis~\cite{lew18} in the late 1920s. 
It garnered significant attention in the 1960s, notably due to contributions from Kripke and Hughes~\cite{kri63,hucr68}. 
Since then, research in this domain has continued to flourish~\cite{go06,bbw06}. 
Modal logic has inspired numerous extensions such as temporal logics~\cite{pr67,pn77,vast85,emju00,clemsi86,sc02}, hybrid logics~\cite{ArBM99b,SchwentickW07,arblma99}, and description logics~\cite{brle84,brsc85,DLHB}.  
Additionally, modal logics have found widespread applications within information and knowledge systems~\cite{eg98,guarino1998formal,longley2015giscience,brewka1997nonmonotonic,DBLP:conf/tark/EngelfrietV98,misiuna2012modal,baader1995terminological}.

The core concept of modal logic is the extension of classical propositional logic through the introduction of two \emph{modalities}: $\Box$ (box) and $\Diamond$ (diamond). 
These modalities facilitate reasoning about necessity (all accessible worlds) and possibility (some accessible worlds). 
Kripke semantics then interpret formulas over frames---labelled states with transition relations---enabling reasoning over sets of worlds rather than single assignments, as in propositional logic.

This extension results in increased computational complexity, with the satisfiability problem being $\PSPACE$-complete. 
Recognising the inherent difficulty, Hemaspaandra~et~al.~\cite{hescsc10} examined the satisfiability problem for modal logic through the lens of Post's lattice~\cite{pos41}, in 2010. 
Informally, their focus was on restricting all possible sets of Boolean functions, in combination with possible subsets of modalities $\{\Box,\Diamond\}$, and classifying these \emph{fragments} of the modal satisfiability problem with respect to different frame classes that stem from the well-known modal logic systems \framefont{K} (all frames), \framefont{KD} (total preorder frames), \framefont{K4} (transitive frames), \framefont{S4} (reflexive and transitive frames), \framefont{S5} (equivalence relation frames, i.e., reflexive, transitive, and symmetric), and \framefont{T} (reflexive frames). 
To streamline notation, we employ the names of the logics to denote their corresponding frame classes. 
Now, Hemaspaandra et~al.~\cite{hescsc10} classified the computational complexity for almost every fragment, except those restricted to the Boolean connectives ${\xor,\true}$---exclusive-or combined with the constant true---over the frame classes \framefont{T}, \framefont{S4}, and \framefont{S5}.
They conjectured that these cases could be solvable in polynomial time, akin to their results for \framefont{K} and \framefont{KD}. 

Interestingly, previous research indicates that problems involving the affine function $\xor$ presents significant challenges. 
Despite several attempts to classify problems related to extensions of Boolean propositional logic using Post's lattice, these fragments have often eluded complete classification (i.e., matching upper and lower complexity bounds)~\cite{bamuscscscvo11,crscthwo10,crscth10,thomas09,bsssv09,MMSTWW09,rei01}. 
Hence, it appears prudent to further examine the impact of $\xor$ on the tractability of such problems.

Therefore, one of our main motivations is to address the counterintuitive nature of the exclusive-or function and the prevailing perplexity it evokes.
The main result of this paper demonstrates that the satisfiability problem for affine multi-modal logic with exclusive-or is NP-hard on S5 frames. 
This result refutes the conjecture of Hemaspaandra~et~al.~\cite{hescsc10} for S5 frames.

The paper is organized as follows. 
First, we introduce the notions of multi-modal logic, Kripke structures, and frame classes. 
\Cref{sec:hard} presents the \NP lower bound proof for the frame class \frameSfive, and we conclude thereafter.

\section{Preliminaries}
\paragraph{Complexity.} In this paper, we use the complexity classes $\P$ and $\NP$. 
The reduction used is a polynomial-time many-one reduction, computable by a deterministic Turing machine running in polynomial time. Further background information can be found in the standard textbook by Papadimitriou~\cite{pap94}.

\paragraph{Modal Logic.} We provide a brief introduction to modal logic; for a more comprehensive treatment, see Blackburn et al.~\cite{bbw06}. 
Let $\phi$ be a propositional formula and $A$ be a set of variables. 
We write $A \models \phi$ if and only if the assignment $\theta(a) = 1$ for all $a \in A$ satisfies $\phi$.

In this paper, we consider $k$-multi-modal formulas over a countably infinite set of propositions $\PROP$, the connective $\oplus$, as well as the unary operators $\Diamond_i$ for $i\in\{1,\dots,k\}$. Formally, the set of all \emph{affine $k$-multi-modal formulas} $\ML{k}{}{}$ is defined by the following grammar:
\[
\varphi \Coloneqq
\true\mid
\false\mid
p\mid
\varphi\xor\varphi\mid
\Diamond_i\varphi,
\]
where $p\in\PROP$. 
As usual, the $\Box$ operator can be defined using negation and $\Diamond$ as follows: $\Box_i\varphi \equiv (\Diamond_i(\varphi \xor \true)) \xor \true$, where $\varphi\xor\true\equiv\lnot\varphi$ is realised. 
We further use $\ML{k}{M}{}$ to denote the set of affine $k$-multi-modal formulas over the modalities in $M \subseteq \{\Box, \Diamond\}$.

From a semantic perspective, we use Kripke structures to define the satisfaction relation $\models$.
\begin{definition}
 Let $M=(W,\mathbf R,\xi)$ be a Kripke structure, i.e., $W$ be a set of worlds, $\mathbf R=\{R_1,\dots,R_n\}$ be a set of transition relations $R_i\colon W\to W$ for $i\in[n]$, and $\xi\colon\PROP\to2^W$ be a labeling function. 
 Furthermore, let $\varphi,\psi\in\ML{k}{\Box,\Diamond}{}$. 
 Then we define for $w\in W$ that
\begin{tabbing}
$M, w \models \varphi \xor \psi$ \= always \kill
$M, w \models \true$ \> always,\\
$M, w \models \false$ \> never,\\
$M, w \models p$ \> if and only if $w \in \xi(p)$,\\
$M, w \models \varphi \xor \psi$ \> if and only if either $M, w \models \varphi$ or $M, w \models \psi$,\\
$M, w \models \Diamond_i \varphi$ \> if and only if there is a $w'$ such that $(w R_i w' \text{ and } M, w' \models \varphi)$.
\end{tabbing}

For formulas in $\ML{1}{\Box,\Diamond}{}$, we simply write $M=(W,R,\xi)$ and omit the bold font $\mathbf R$ to indicate that we consider only a single transition relation. 
We say that a formula $\varphi$ is \emph{satisfiable} if and only if there exists a Kripke structure $M=(W,\mathbf R,\xi)$ and a world $w\in W$ such that $M,w\models\varphi$ is true. 
Furthermore, we call $(W,\mathbf R)$---that is, a model without a labeling function $\xi$---a \emph{frame}.
\end{definition}

In this paper, we consider frames that satisfy specific properties, such as transitivity or reflexivity. 
In particular, we work with the frame classes shown in \Cref{fig:frames}. 
If a given Kripke structure $M$, or a frame $(W,\mathbf R)$, satisfies the properties of some frame class $\framefont F$, then we say $M$ is an $\framefont F$-structure, or $(W,\mathbf R)$ is an $\framefont F$-frame.

\begin{figure}[t]
\begin{center}
$\begin{array}{cp{.5em}l}\toprule
\textbf{Frame class} && \textbf{ Description}\\\midrule
 \frameT && \text{ the set of reflexive frames}\\
 \frameSfour && \text{ the set of reflexive, transitive frames}\\
 \frameSfive && \text{ the set of reflexive, transitive, symmetric frames}\\\bottomrule
\end{array}$\\[1em]
$\begin{array}{lp{.5em}l}\toprule
\textbf{Frame property} && \textbf{FO definition}\\\midrule
\text{reflexive frames} &&\forall w\in W: wRw\\
\text{transitive frames} &&\forall u,v,w\in W : (uRv\land vRw)\to uRw\\
\text{symmetric frames} &&\forall u,v\in W: uRv\to vRu\\\bottomrule
\end{array}$
\end{center}
\caption{Frame classes and First-Order definitions of frame properties.}\label{fig:frames} 
\end{figure}

The decision problem of interest is the satisfiability problem for affine $k$-multi-modal logic. The general definition of this problem is as follows.
\begin{framed}\vspace{-.4cm}
\begin{description}[style=nextline, align=right, labelwidth=2cm, labelindent=-.75em, leftmargin=!]
 \item[Problem:] $\SAT{\framefont F}{k}{M}{}$
 \item[Input:] A formula $\varphi\in\ML{k}{M}{}$.
 \item[Question:] Does there exist an $\framefont F$-structure $M=(W,\mathbf R,\xi)$, and a world $w\in W$ such that $M,w\models\varphi$?
\end{description}\vspace{-.4cm}
\end{framed}
In the case we investigate, we have $M = \{\Box, \Diamond\}$.
Given a Kripke structure $M = (W, \mathbf{R}, \xi)$, we define a \emph{substructure} $S \subseteq M$ to be of the form $S = (W', \mathbf{R}|_{W'}, \xi|_{W'})$, where $W' \subseteq W$, $\mathbf{R}|_{W'}$ is the restriction of each relation $R \in \mathbf{R}$ to $R \cap (W' \times W')$, and $\xi|_{W'}$ is the restriction of $\xi$ to worlds in $W'$.

\section{Affine ML-SAT on S5 Frames is NP-Hard}\label{sec:hard}
In this section, we are going to prove the \NP-hardness result of our problem with respect to the frame class \frameSfive. 

Consider a 3CNF formula $f=\bigwedge_{i=1}^l C_i$, where $C_i$ is a clause with 3 literals, i.e., $C_i=l_{i1}\lor l_{i2}\lor l_{i3}$. The literals are either $a$ or $\lnot a$ for variables $a\in A$, and $A$ is the set of variables.

\begin{remark}\label{bracketing}
We assume that the parentheses are inserted such that the connective $\land$ only appears as a binary connective and the depth of the formula is logarithmic in the number of clauses. So every subformula of $f$ is either of the form $(l_{i1}\lor l_{i2}\lor l_{i3})$ or is a conjunction $(g\land h)$, where $g$ and $h$ are subformulas of $f$. This will be a crucial point in the course of the proof.    
\end{remark}

Let us now define the property of independent subformulas, which will be used later in the proof of \Cref{thm:np-hard}.
\begin{definition}\label{def:independent}
Let $g,h$ be subformulas of $f$. We say that $g$ is \emph{independent} from $h$ if there exists a clause in $g$ consisting of variables which all are not present in $h$. 
If $g$ is independent from $h$ and $h$ is independent from $g$ then we say $g$ and $h$ are \emph{strongly independent}.
\end{definition}
\begin{example}
 Let $f=(a\lor b\lor c) \land (\overline d\lor f\lor e)\land(\overline e\lor f \lor g)$, and the subformulas $g=(a\lor b\lor c)$, $h=(\overline d\lor f\lor e)\land(\overline e\lor f\lor g)$. Then $g$ and $h$ are strongly independent, but $g'=g\lor(\overline d\lor f\lor e)$ is not strongly independent of $h$.
\end{example}

This property will ensure that in our upcoming construction we never have a subformula $(g\land h)$ where $g \equiv h$.

\begin{theorem}\label{thm:np-hard}
 $\SAT{\frameSfive}{2}{\Box,\Diamond}{}$ is \NP-hard.
\end{theorem}
\begin{customproof}{Thm.~\ref{thm:np-hard}}
 The main idea is a reduction from $\threeSAT$ to $\SAT{\frameSfive}{2}{\Box,\Diamond}{}$.
Given a 3CNF formula, we transform it into another 3CNF formula such that, for every subformula $g \land h$, $g$ and $h$ are independent, while preserving the satisfiability of the original formula.

\begin{lemma}\label{lem:independent}
Let $A$ and $B$ be two disjunct sets of variables. 
Given a formula $f=\bigwedge_{i=1}^l C_i$ over the variables $A$ then the formula $f'=\bigwedge_{i=1}^l (C_i \land (b_i\lor b_i\lor b_i))$ over the variables $B=\{b_1,\dots,b_l\}$ is satisfiable iff $f$ is satisfiable.
Moreover for any subformula $(g\land h)$ of $f'$, we have that $g$ and $h$ are strongly independent.
\end{lemma}
\begin{customproof}{Lem.~\ref{lem:independent}}
It is clear that $f$ is satisfiable iff $f'$ is satisfiable, since when all variables of $B$ are assigned to true $f$ and $f'$ are syntactically the same formula.

Also given any subformula $(g\land h)$ of $f'$, then either $h=(b_i\lor b_i\lor b_i)$ for some $i$ or both $g$ and $h$ contain a clause consisting of a different variable of $B$.
\end{customproof}

Using the previous lemma, we will assume from now on that for every subformula $g \land h$ of $f$, $g$ and $h$ are independent.

We are now able to define the inductive translation $\phi_{(\cdot)}$ from propositional 3CNF formulas to formulas in $\ML{2}{\Box,\Diamond}{}$ as follows:
\begin{align*}
 \phi_a &=(\square_1 a)\oplus(\square_2\square_1 a) &&,\text{ if $a$ is a positive literal}\\
 \phi_{\neg a}&=(\square_1 \neg a)\oplus(\square_2\square_1 \neg a) &&, \text{ if $\lnot a$ is a negative literal},\\
 \phi_{C}&=\Diamond_1(\phi_{l_1}\oplus\phi_{l_2}\oplus\phi_{l_3})&&,\text{ if $C=l_1\lor l_2\lor l_3$ is a clause},
\end{align*}
and given a subformula $(g\land h)$ we define
\begin{align*}
\psi_{g}&=\phi_{g}\oplus\square_2\phi_g,\quad\text{ and }\quad\psi_{h}=\phi_{h}\oplus\square_2\phi_h,\\
\phi_{g\land h}&=(\Diamond_1(\psi_g\oplus\psi_h))\oplus(\Diamond_1\psi_g)\oplus(\Diamond_1\psi_h).
\end{align*}
With an appropriate bracketing of $f$ (see \Cref{bracketing}), the formula $\phi_f$ will have polynomial size in the size of $f$.
Note that $\psi_g$ and $\psi_h$ are satisfied in subparts of any model, whereas $\psi_{g\land h}$ is either true or false in the complete model.

We now claim that $f$ is satisfiable if and only if $\phi_f$ is satisfiable.
We will prove this claim in two separate statements. First, consider the easy direction~``$\Rightarrow$''.

\begin{proposition}\label{prop:easy}
If $\phi_f$ is satisfiable, then $f$ is satisfiable.
\end{proposition}
\begin{customproof}{Prop.~\ref{prop:easy}}
Assume that $\phi_f$ is satisfiable, i.e., there exists $M,w$ such that $M,w\models\phi_f$. Consider the set $A = \{\,a_i \mid M,w \models a_i\,\}$.
We show that $A \models f$ by induction on the structure of $f$.
Let $W=\{\,w'\mid w\sim_1 w'\,\}$. 
Our inductive hypothesis is if there exists a state $w' \in W$ such that $W, w' \models \phi_f$ then $A \models f$.

Let $f'$ be an arbitrary subformula of $f$.
\begin{description}
\item[Case $f' = a$:]

Let $w' \in W$ such that $w' \models \phi_a$. Then $M,w'\models (\square_1 a)\oplus(\square_2\square_1 a)$ (by definition of $\phi_a$), which implies that $M,w'\models (\square_1 a)$ (by reflexivity of $M$), and hence for every $w''\in W$ we have that $M,w''\models a$ (as all worlds in $W$ are $\sim_1$-reachable) and therefore $a \in A$. 

 By a symmetric argument $M,w''\models\phi_{\neg a}$ implies for every $w'''\in W$ we have that $M,w'''\models \neg a$. Hence, if there exist $w'\in W$ such that $w'\models\phi_{\neg a}$ then $a\notin A$.
 
\item[Case $f' = l_1 \lor l_2 \lor l_3$:]

Firstly, note that if there exists a $w'\in W$ such that $M,w'\models\phi_{l_1\lor l_2\lor l_3}$, then there exists $w''\in W$ such that $M,w''\models\phi_{l_i}$ for some $i=1,2,3$. Then, by induction hypothesis, we know that $A \models l_i$. Therefore, we have that $A \models l_1 \lor l_2\lor l_3$.

\item[Case $f' = g \wedge h$:]

We show that if there exists a $w'\in W$ such that $M,w'\models\phi_{g\land h}$, then there exists $w'',w'''\in W$ such that $M,w''\models\phi_{g}$ and $M,w'''\models\phi_{h}$.
Assume, w.l.o.g.\ that for all $w''$, $M, w'' \not \models \phi_g$. Therefore for all $w''$, we have $M, w'' \not\models \psi_g$ and hence $M, w'' \not \models \phi_{g \wedge h}$.

So we assume that there exists $w'', w''' \in M$, such that $M, w'' \models \phi_g$ and $M, w''' \models \phi_h$. By induction hypothesis, this implies $A \models g$ and $A \models h$. Therefore, we have that $A \models g \wedge h$.
\end{description}
Hence, we have that $A \models f$ and the proposition holds.
\end{customproof}

Now consider the more difficult part to prove, i.e., the direction ``$\Leftarrow$''.
\begin{proposition}\label{prop:difficult}
 If $f$ is satisfiable then $\phi_f$ is satisfiable.
\end{proposition}
\begin{customproof}{Prop.~\ref{prop:difficult}}
Fix a formula $f$, and a satisfying assignment $A$.
For a model $M$ and an assignment $A$, we define the $A$-\emph{core} of $M$ to be a nonempty set $S \subseteq M$ which satisfies the following properties:
\begin{itemize}
\item $S$ is an equivalence class under $\sim_1$, and
\item for all $w\in S$ we have that $M,w\models a_i$ if and only if $A\models a_i$.
\end{itemize}

Let $g$ be a subformula of $f$. We say that a model $M$ is a \emph{nice} model for $g$ if $M$ has a non-empty core $S$ in which all worlds satisfy $g$.
Looking back at the proof of \Cref{prop:easy}, one will notice that we actually proved that every satisfying model for $\phi_f$ is, in fact, a nice model for $f$. For this direction, we will construct nice models by induction over the structure of the formula, starting with single clauses and providing a construction for $(g\land h)$.

\begin{lemma}
\label{lem_base}
For every clause $C$, we have a nice model for the formula $\phi_{C}$.
\end{lemma}
\begin{customproof}{Lem.~\ref{lem_base}}
Let $C = l_1 \vee l_2 \vee l_3$ be some clause and $A$ be a valid assignment for $f$. 
Therefore, we know that $A \models C$, that is, one of the literals, say $l_j$, for a $j \in [3]$ is satisfied by $A$. Consider an assignment $A'$ such that $A' \models l_{j'}$ for all $j' \neq j$ and $A' \not \models l_j$. Of course, such an assignment exists.

Now define the nice model as depicted in \Cref{fig:lem_base}. The model $M$ for $\phi_{C}$ consists of states $w$ and $v$, where the propositions in $w$ is $A$ and the propositions in $v$ is $A'$. We have the following relations among the states: $(w \sim_{\{1,2\}} w), (v \sim_{\{1,2\}} v), (w \sim_2 v)$.

Then it is clear that $M, w \models \phi_{l_{j'}}$ and for all $j' \neq j$, we have that $M, w \not \models l_j$. Therefore, we have that $M, w \models \phi_{C_i}$.

Note that the core of this model consists of just the single state $w$. Also observe that all the properties required for a nice model are satisfied.
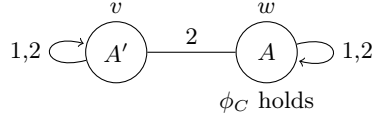
\begin{figure}[t]
 \centering
 \begin{tikzpicture}
  \node[state,label={90:$v$}] (v) at (0,0) {$A'$};
  \node[state,label={90:$w$},label={-90:$\phi_C$ holds}] (w) at (2,0) {$A$};
 
  \path (v) edge [above] node {2} (w)
  (v) edge [loop left] node {1,2} (v)
  (w) edge [loop right] node {1,2} (w);
 \end{tikzpicture}
\caption{Satisfying nice model for \Cref{lem_base}.}\label{fig:lem_base}
\end{figure}
\end{customproof}

For the subsequent construction, we need to find a model for $(g\land h)$ given models for $g$ and $h$. The following result shows that the nice property of models is preserved under arbitrary extensions.

\begin{lemma}
\label{lem_merge}
Let $g$ be a formula and let $M$ be a nice model for $g$ with core $S$. Let $M'$ be an arbitrary model with (a possibly empty) core $S'$. Then there exists a nice model $\hat M$ which satisfies $g$.
\end{lemma}
\begin{customproof}{Lem.~\ref{lem_merge}}
Consider the new model $\hat M = M \cup M'$, where $S \cup S'$ are pairwise connected by $\sim_1$. Then $\hat M$ is a nice model with core $S \cup S'$. In particular, for all $m \in S \cup S'$, we have that
\[\hat M, m \models \phi_g.\]
Due to the form of $\phi_g$, which begins with a $\Diamond_1$ operator, this property holds immediately.
\end{customproof}

The next lemma demonstrates an interesting property of \emph{independent} formulas (see \Cref{def:independent}). Recall that we assume for any subformula $(g \land h)$ of $f$, the formulas $g$ and $h$ are independent.

\begin{lemma}
\label{lem_indepForm}
Let $g,h$ be independent subformulas of $f$, and let $M$ be model and world $m \in M$ such that $M,m \models \phi_g \wedge \phi_h$. Then there exists a model $N$ and world $n \in  N$ such that $N,n \models \phi_g \wedge \neg \phi_h$.
\end{lemma}
\begin{customproof}{Lem.~\ref{lem_indepForm}}
This follows directly from the definition of independence: since we can find a clause in $h$ whose variables do not appear in $g$, we can falsify this clause and thereby falsify $h$, while the satisfiability of $g$ remains unaffected.
\end{customproof}

We now claim that for any formula $\phi_{h \wedge g}$, where $h$ and $g$ are independent and both are conjunctions of clauses, we can construct a nice model.

\begin{lemma}
\label{lem_ih}
Let $g$ and $h$ be independent subformulas of $f$ that correspond to sets of clauses. If there are nice models for $\phi_h,\phi_g$, then there exists a nice model for $\phi_{h\land g}$.
\end{lemma}
\begin{customproof}{Lem.~\ref{lem_ih}}
Let $M_g,M_h$ be nice models of $\phi_g,\phi_h$ with cores $S_g,S_h$, respectively. Furthermore, let $w_g, w_h$ be two additional worlds with propositions $A$ (recall that $A$ is the satisfying assignment of $f$). 
We connect all worlds of $S=\{w_g,w_h\}\cup S_g\cup S_h$ pairwise via $\sim_1$ giving the set $S$.

Applying Lemma \ref{lem_indepForm} on copies of $M_g$ and $M_h$ then constructs models $\bar M_g$, $\bar M_h$, and worlds $v_g$ and $v_h$ such that $\bar M_g, v_g \models \phi_g \wedge \neg \phi_h$ and $\bar M_h, v_h \models \phi_h \wedge \neg \phi_g$.
We now make a $\sim_2$-edge between $w_g$ and $v_h$ and a $\sim_2$-edge between $w_h$ and $v_g$ as shown in Figure \ref{fig_merge}.

\begin{figure}
\begin{center}
\begin{tikzpicture}[every node/.style={draw,inner sep=0.5mm},scale=0.6,transform shape,font=\Large]
\node[star, star points=40, star point ratio=0.98,minimum height=4.8cm,label={[label distance=-1cm]160:$M_g$}]  (M1) at (-3,3)  {}; 
\node[star, star points=40, star point ratio=0.98,minimum height=4.8cm,label={[label distance=-1cm]20:$M_h$}]  (M1) at (3,3)  {};
\node[star, star points=40, star point ratio=0.97,minimum height=3.2cm,label={[label distance=-1cm]160:$S_g$}]  (M1) at (-2.6,2.4)  {};
\node[star, star points=40, star point ratio=0.97,minimum height=3.2cm,label={[label distance=-1cm]20:$S_h$}]  (M1) at (2.6,2.4)  {};
\node[circle] (S11) at (-3.2,1.8) {$u_1$};
\node[circle] (S12) at (-2.5,3.2) {$u_2$};
\node[circle] (S13) at (-1.8,2) {$u_3$};
\node[circle] (S21) at (3.2,1.8) {$v_1$};
\node[circle] (S22) at (2.5,3.2) {$v_2$};
\node[circle] (S23) at (1.8,2) {$v_3$};

\node[circle] (SG) at (-1,0) {$w_g$};
\node[circle] (SH) at (1,0) {$w_h$};

\node[star, star points=40, star point ratio=0.98,minimum height=4.8cm,label={[label distance=-1cm]200:$M'_h$}]  (MXG) at (-3,-3)  {};
\node[star, star points=40, star point ratio=0.98,minimum height=4.8cm,label={[label distance=-1cm]340:$M'_g$}]  (MXH) at (3,-3)  {};
\node[star, star points=40, star point ratio=0.97,minimum height=3.2cm,label={[label distance=-1cm]200:$S'_h$}]  (SXG) at (-2.6,-2.4)  {};
\node[star, star points=40, star point ratio=0.97,minimum height=3.2cm,label={[label distance=-1cm]340:$S'_g$}]  (SXH) at (2.6,-2.4)  {};
\node[circle] (SX11) at (-3.2,-1.8) {$v'_1$};
\node[circle] (SX12) at (-2.5,-3.2) {$v'_2$};
\node[circle] (SX13) at (-1.8,-2) {$v'_3$};
\node[circle] (SX21) at (3.2,-1.8) {$u'_1$};
\node[circle] (SX22) at (2.5,-3.2) {$u'_2$};
\node[circle] (SX23) at (1.8,-2) {$u'_3$};

\path[draw]  (S11) -- (S12) -- (S13) -- (S11);
\path[draw]  (S21) -- (S22) -- (S23) -- (S21);
\path[draw]  (SX11) -- (SX12) -- (SX13) -- (SX11);
\path[draw]  (SX21) -- (SX22) -- (SX23) -- (SX21);

\foreach \from/\to in {SG/S11, SG/S13, SG/S21, SG/S23, SH/S11, SH/S13, SH/S21, SH/S23, S12/S22, S13/S23, S12/S23, S13/S22}
 \path[draw=black, thick] (\from) -- (\to); 

\path[draw,dashed,thick](SG) --   node[draw=none,above,very near end] {$2$} (SX11);
\path[draw,dashed,thick](SH) --  node[draw=none,above,very near end] {$2$} (SX21);
\node[ellipse,minimum height=6cm, minimum width=9cm, fill, fill opacity=0.1, draw=none] at (0,2) {};
\node[draw=none] at (0,4.5) {$S$};

\end{tikzpicture}

\end{center}
\caption{A nice model for conjunction used in the proof of \Cref{lem_ih}. Thick lines are $\sim_1$ transitions, wherease thick dashed lines are $\sim_2$ transitions.}
\label{fig_merge}
\end{figure}
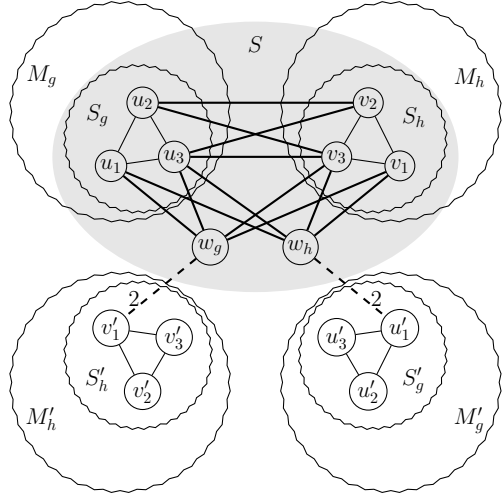

Now we claim that the resulting model $M$ is a nice model with core $S$.

Due to Lemma \ref{lem_merge} the following is true: For all $w \in S$, we have that $M, w \models \phi_g$, and for all $w \in S$, we have that $M, w \models \phi_h$.

By the additional $\sim_2$-edges at $w_g,w_h$ which are the only (!) $\sim_2$-edges at these states we have that $M, w_g \models \psi_g$ and $M, w_h \models \psi_h$. 
Then also $M, w_h \models \neg\psi_g$ and $M, w_g \models \neg\psi_h$, as in $v_g$ and in $w_h$ the formula $\phi_g$ is true.

The last fact immediately gives the following.  
For all $w \in S$, we have that $M, w \models \Diamond_1 \psi_g$. 
For all $w \in S$, we have that $M, w \models \Diamond_1 \psi_h$. 
For all $w \in S$, we have that $M, w \models \Diamond_1 (\psi_g\oplus\psi_h)$, due to $w_g$ (resp., $w_h$).

Hence, for all $w \in S$, we have that $M, w \models \phi_{g \wedge h}$.
\end{customproof}

Finally, we can complete the proof of \Cref{prop:difficult}, showing that if $f$ is satisfiable, then $\phi_f$ is satisfiable. 

Assume that $f$ is satisfiable. We now show, by induction on the depth of the formula, that there exists a model $M_f$ and a state $m_f$ such that $M_f, m_f \models \phi_f$.

The base case is given by Lemma \ref{lem_base}. Every clause $C_i$ is now a conjunction with the $B_i$ propositional clause. The clauses $C_i$ and $B_i$ satisfy the condition for independent subformulas and therefore we can apply Lemma \ref{lem_ih} and get a nice model which satisfies $\phi_{C_i \wedge b_i}$. We can continue applying Lemma \ref{lem_ih}, since there will always be a literal different in the combining subformulas.
\end{customproof}
Since \Cref{prop:easy,prop:difficult} have been proven and the reduction can be computed in polynomial time, the theorem follows.
\end{customproof}

\subsection*{A few insights into the two left open cases}

As noted in the introduction, two cases from Hemaspaandra~et~al.~\cite{hescsc10} remain open: $\SAT{\frameSfour}{1}{\Box,\Diamond}{}$ and $\SAT{\frameT}{2}{\Box}{}$.

We are still investigating these cases but want to offer preliminary insights here. The NP-hardness proof above relied crucially on frame symmetry to construct nice models, a property absent in $\frameSfour$ and $\frameT$. This precludes reusing cores as before, rendering such constructions inapplicable.

Now consider $\phi\xor\psi$ where $\phi$ and $\psi$ have unequal modal depths. For satisfiability, a model of sufficient depth can falsify the deeper formula outside the submodel evaluating the shallower one. This observation could help in adapting the polynomial-time algorithm of Hemaspaandra~et~al.~\cite{hescsc10} to these cases, supporting their conjecture. Yet, we have not fully formalised this intuition into a complete and rigorous proof.

\section{Conclusion}
In this paper, we have resolved one of the open cases identified by Hemaspaandra~et~al.~\cite{hescsc10}. 
Specifically, our result demonstrates that, for the frame class $\frameSfive$, their conjecture does not hold under the widely accepted complexity assumption that $\P\neq\NP$. 
In particular, we have established the $\NP$-hardness of the satisfiability problem in this setting.

An intriguing observation emerges regarding the role of symmetry in $\frameSfive$, which appears to be the pivotal factor enabling this $\NP$-hardness. 
In reflexive and transitive frames, the modalities allow movement only forward through the model, without the ability to refer back to previous states. 
Consequently, interactions between different chains of states seem impossible. 
However, the presence of symmetry changes this dynamic, as it permits returning to earlier states. This capability enforces contradictions in regions of the model that were visited previously, thereby facilitating the hardness construction.

As future work, we plan to tackle the two remaining open cases from \cite{hescsc10}, namely those concerning the frame classes $\frameSfour$ and $\frameT$. 
Additionally, further research could explore the isolated influence of symmetry by examining the complexity of the modal satisfiability problem in purely reflexive and symmetric frames. 
The other open cases for different logics mentioned in the introduction also present promising avenues for investigation.

\begin{credits}
\subsubsection{\ackname} The second author appreciates funding by the German Research Foundation (DFG) under the grant ME~4279/3-1.

\subsubsection{\discintname}
The authors have no competing interests to declare that are
relevant to the content of this article.

\end{credits}

\bibliographystyle{splncs04}
\bibliography{mlsats5-refs}
\end{document}